\documentclass[sigconf]{acmart}
\AtBeginDocument{%
  \providecommand\BibTeX{{%
    \normalfont B\kern-0.5em{\scshape i\kern-0.25em b}\kern-0.8em\TeX}}}

\copyrightyear{2021} 
\acmYear{2021} 
\setcopyright{acmcopyright}\acmConference[ESEM '21]{ACM / IEEE International Symposium on Empirical Software Engineering and Measurement (ESEM)}{October 11--15, 2021}{Bari, Italy}
\acmBooktitle{ACM / IEEE International Symposium on Empirical Software Engineering and Measurement (ESEM) (ESEM '21), October 11--15, 2021, Bari, Italy}
\acmPrice{15.00}
\acmDOI{10.1145/3475716.3475776}
\acmISBN{978-1-4503-8665-4/21/10}

\usepackage{fontawesome}

\usepackage{tcolorbox}
\definecolor{myBlue}{RGB}{240,220,255}
\usepackage{soul}
\usepackage{color}

\begin{document}

\title{An Empirical Analysis of Practitioners' Perspectives on \\ Security Tool Integration into DevOps}

\author{Roshan Namal Rajapakse}
\affiliation{%
  \institution{The University of Adelaide}
   \institution{Cyber Security Cooperative Research Centre}
  \country{Australia}}
\email{roshan.rajapakse@adelaide.edu.au}

\author{Mansooreh Zahedi}
\affiliation{%
  \institution{The University of Adelaide}
  \country{Australia}}
\email{mansooreh.zahedi@adelaide.edu.au}

\author{Muhammad Ali Babar}
\affiliation{%
  \institution{The University of Adelaide}
   \institution{Cyber Security Cooperative Research Centre}
  \country{Australia}}
\email{ali.babar@adelaide.edu.au}

\renewcommand{\shortauthors}{Rajapakse et al.}

\begin{abstract}

\textbf{Background:} Security tools play a vital role in enabling developers to build secure software. However, it can be quite challenging to introduce and fully leverage security tools without affecting the speed or frequency of deployments in the DevOps paradigm. \textbf{Aims:} We aim to empirically investigate the key challenges practitioners face when integrating security tools into a DevOps workflow in order to provide recommendations to overcome them. \textbf{Method:} We conducted a study involving 31 systematically selected webinars on integrating security tools in DevOps. We used a qualitative data analysis method, i.e., thematic analysis, to identify the challenges and emerging solutions related to integrating security tools in rapid deployment environments. \textbf{Results:} We find that while traditional security tools are unable to cater for the needs of DevOps, the industry is moving towards new generations of tools that have started focusing on these requirements. We have developed a DevOps workflow that integrates security tools and a set of guidelines by synthesizing practitioners' recommendations in the analyzed webinars. \textbf{Conclusion:} While the latest security tools are addressing some of the requirements of DevOps, there are many tool-related drawbacks yet to be adequately addressed.

\end{abstract}




\maketitle

\section{Introduction}

DevOps (Software development (Dev) and information technology operations (Ops)) paradigm is aimed at removing the traditional boundaries (or \textit{silos}) between software development and operations teams \cite{jabbari2016devops, lwakatare2016exploratory, ebert2016devops}. The main intended goal of this paradigm is to enable organizations to react more quickly and flexibly to changes in the business environment or observed behaviours of the system  {\cite{sharma2015devops}}. The elimination of the separation between development and operations activities enables achieving this goal while ensuring high quality of the output.

However, with the gains in speed, practitioners are reportedly facing several challenges in delivering secure software in the DevOps paradigm \cite{rahman2016software,rafi2020prioritization}.  In the traditional software development paradigms, security practices such as Static Application Security Testing (SAST) \cite{yang2019towards}, Dynamic Application Security Testing (DAST) \cite{peterson2020} are carried out at a later stage of the cycle. To implement these practices, practitioners use \textit{application security testing tools}, which test applications for security vulnerabilities \cite{Gartner}. However, these tools have several limitations that make them unsuitable for security testing in DevOps. \cite{rajapakse2021challenges}. For example, traditional security testing tools frequently produce inaccurate results (e.g., false positives and negatives) \cite{chess2004static, johnson2013don}, which requires an engineer to manually assess the accuracy of these outputs. Such manual processes are discouraged in DevOps, with \textit{automation} frequently cited as a key pillar in this paradigm \cite{tomas2019empirical}. Recent industry reports also conclude that the shortcomings of security tools such as poor quality of scanning can harm developers' productivity \cite{Shiftleft2020}. Accordingly, practitioners are finding it challenging to conduct those security practices in DevOps (e.g., SAST and DAST) due to security tool limitations.

Our main objective in this study is to empirically investigate how security tools can be integrated into a DevOps workflow without affecting its rapid delivery goals. We use the term \textit{security tools} to address both application security testing tools (e.g., SAST and DAST) and run-time security tools, which are used to prevent exploits (e.g., Web Application Firewalls (WAF) \cite{razzaq2013critical}). A \textit{workflow} in this context refers to the continuous sequence of software development activities used in DevOps to produce outputs \cite{ben2021devops}. 

To achieve this aim, we have systematically selected and analyzed practitioner discussions on security tool adoption in DevOps, captured as webinars. We include 31 such webinars that contain a rich amount of practice-based information and qualitative data in our study. Having used the thematic analysis method for the analysis of this data, we provide the following key contributions.  

\begin{itemize}
    \item We identify the key challenges practitioners face in integrating security tools into the DevOps workflow.
    \item We present the current and emerging security tools and  recommendations for using such tools in a DevOps environment.
    \item Based on the synthesis of the solutions reported in the webinars, we present a \textit{DevOps workflow} that integrates the new generation of security tools and their usage recommendations.
\end{itemize}

\section{Related work}

\subsection{Security in DevOps}
Security in DevOps (or \textit{DevSecOps}) implies integration of security principles and controls (i.e., processes, tools, and methods) in the DevOps processes \cite{myrbakken2017devsecops, mohan2016secdevops}. Most importantly, it advocates frequent and continuous collaboration between development, operations, and security teams \cite{rahman2016software, myrbakken2017devsecops}. Due to the rapid delivery needs, developers are required to do certain security assessments while engaging in the development activities \cite{sanchez2018characterizing}. 

Rahman and Williams \cite{rahman2016software} summarized the experiences in utilizing security practices in DevOps, using an analysis of internet artifacts and a practitioner survey. They report that software practitioners have mixed opinions about automated deployments in DevOps. One reason for this was the security concerns related to using improper automated deployment tools in DevOps \cite{rahman2016software}.

Rafi et al. \cite{rafi2020prioritization} report a study that extracted the security challenges in DevOps from the literature and evaluated them using a survey. One of their conclusions is that the \textit{lack of automated testing tools} is the most critical challenge to secure DevOps implementations. Accordingly, they recommended that proper testing tools need to be in place to monitor the security risks of DevOps \cite{rafi2020prioritization}.  

Mohan and Othmane \cite{mohan2016secdevops} conducted a mapping study that has identified a set of tools useful in a DevOps environment. However, they do not report how these tools can be adopted into the DevOps workflow or details related to integration problems. 

The review studies by Myrbakken and Colomo-Palacios \cite{myrbakken2017devsecops}, Mao et al. \cite{mao2020preliminary}, and Rajapakse et al. \cite{rajapakse2021challenges} report security tool-related limitations in DevSecOps. Myrbakken and Colomo-Palacios \cite{myrbakken2017devsecops} state that if the security functionality is not automated in the available tools, this would result in friction in the DevOps cycle. They report the studies that have described the difficulties faced by developers in producing secure code in DevOps while using these tools. Mao et al. \cite{mao2020preliminary} have identified the lack of DevSecOps tools and mature DevSecOps solutions as two main implementation challenges in this area. They also state that the complexity of integrating security tools is an issue practitioners need to consider, especially if there is a limited budget for the DevSecOps implementation. Rajapakse et al. \cite{rajapakse2021challenges} have also identified the challenges of integrating traditional security tools in DevOps projects. 

Based on the above studies (e.g., \cite{myrbakken2017devsecops}, \cite{mao2020preliminary}, \cite{rajapakse2021challenges}) there is a need for more research directed at the security tool integration problem in DevOps. Our study aims to target this specific gap in the literature. Also, compared with the previous studies (e.g., \cite{mohan2016secdevops}), we provide a more in-depth analysis of the practice-based information related to the current and emerging security tools landscape. 

\subsection{Security tool adoption}

While there is an extensive range of security tools available, developers do not always use them \cite{xiao2014social}. To understand the reasons behind this issue, Witschey et al. \cite {witschey2014technical} have developed a theoretical model of factors that influence developers’ security tool adoption decisions. This model includes 14 factors divided into four categories: Innovation, Social System, Communication Channel, and Potential Adopter. The authors have found that the developers’ ability to observe their peers using security tools was the strongest predictor for such tool usage (e.g., Social System). Xiao et al. \cite{xiao2014social} have also reported how security tool adoption depends on developers’ social environments and the medium in which they received the information. To motivate security workers to adopt and use security tools, Jordan et al. \cite{jordan2014designing} have developed a system that used persuasive techniques (e.g., automated emails for developers). Our work differs from these studies by focusing on security tool adoption into a DevOps environment, which has a different set of demands.

Concerning the adoption of specific tools, the previous studies have focused on limitations of static \cite{johnson2013don, tomasdottir2018adoption} and dynamic analysis \cite{gosain2015survey} tools in the traditional setup. The specific focus of our study is to understand how these drawbacks affect security tool integration in the DevOps workflow based on practitioner feedback. Also, none of the previous studies in this area has evaluated how current and emerging security tools can be successfully (i.e., without affecting deployment frequency) integrated into the DevOps workflow to the best of our knowledge.

\begin{figure*}
    \centering
    \includegraphics[origin=c, width=1\textwidth]{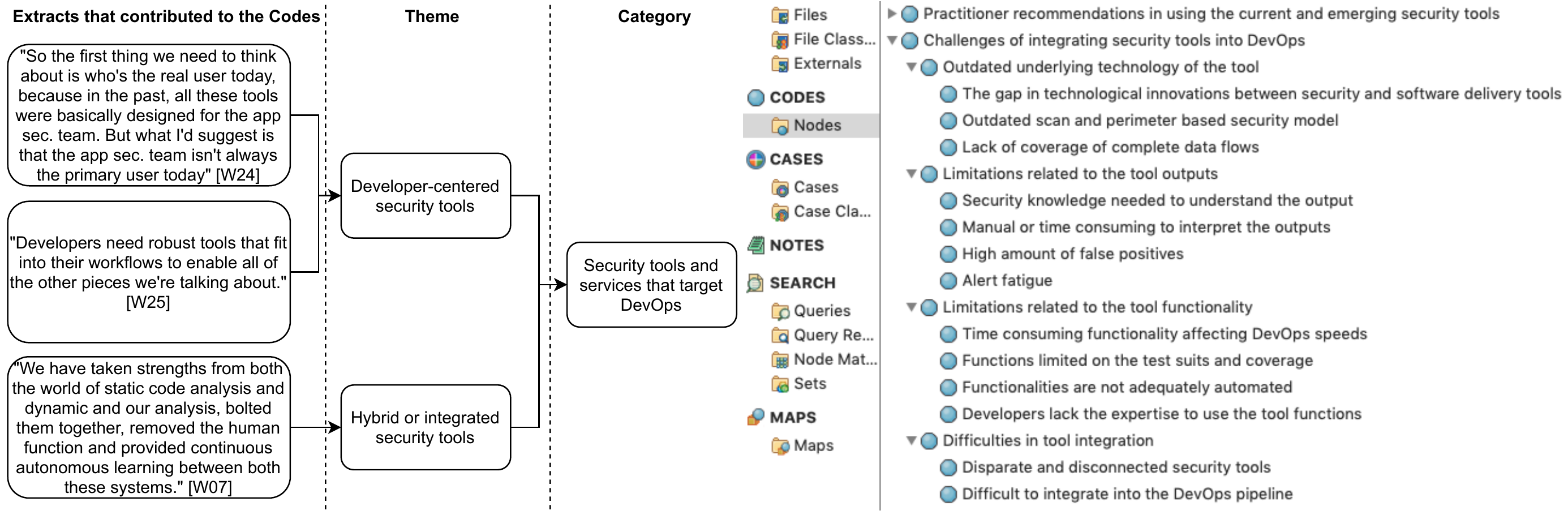}
    \vspace{-6mm}
    \caption{A) An example of the coding process \hspace{5mm} B) NVivo coding structure [Category-Theme-Code]}
    \label{fig:coding}
\end{figure*}

\begin{figure*}[t]
    \centering
    \includegraphics[origin=c, width=.90\textwidth]{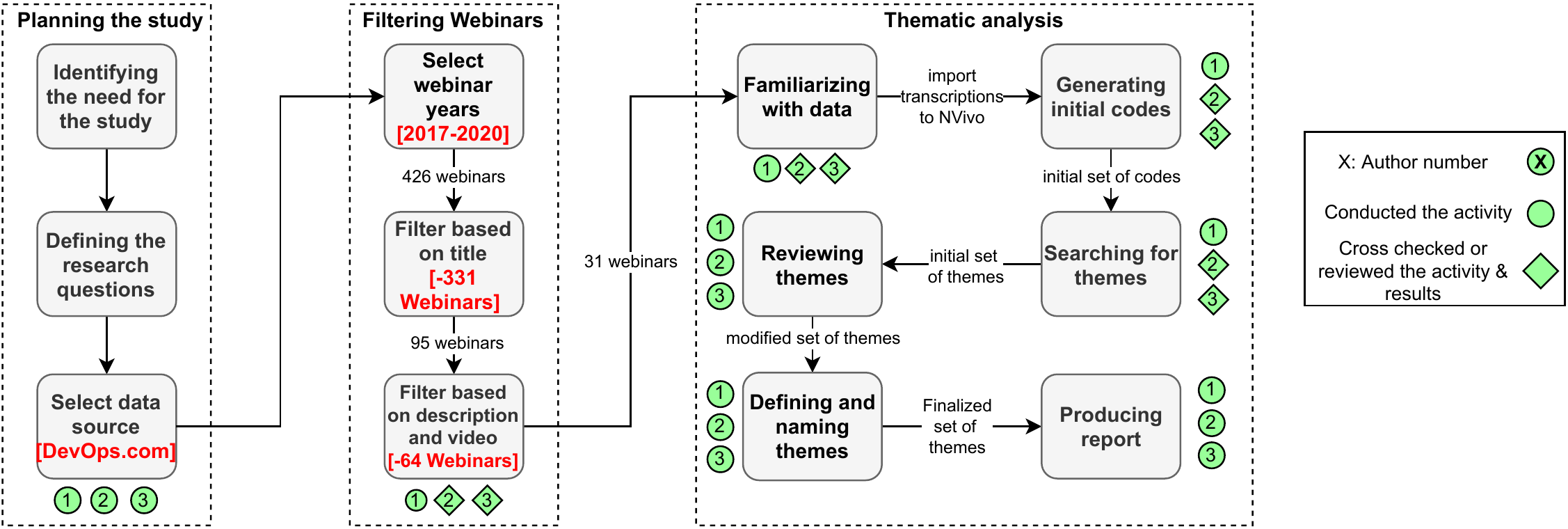}
    \vspace{-2mm}
    \caption{The overview of the method}
    \label{fig:method}
\end{figure*}

\section{Research method}
To address the gap areas in the literature noted above, we investigate the following two research questions in this study:

\textbf{RQ1}: What are the key challenges of integrating security tools into the DevOps workflow? 

\textbf{RQ2}: What are the practitioner recommendations to integrate the current and emerging security tools in a DevOps workflow?

\subsection{Data source}
We conducted this study using webinars in which practitioners discuss the contemporary problems, solutions and emerging trends of integrating security tools in DevOps. We used \textit{DevOpsTV}, the YouTube channel of DevOps.com\footnote{https://devops.com/on-demand-webinars/}, a popular source of information (e.g., webinars) for practitioners in this area as our source. A webinar is generally one hour long and consists of practitioners discussing a timely topic of interest for the DevOps community. These practitioners were either leading DevOps experts or representatives of companies that produce DevOps solutions. Therefore, these resource persons were able to describe the challenges in this domain in detail. Further, they presented comparative analysis on the current and emerging solutions, often accompanied by tool demonstrations. Therefore, we decided that this data source was suited to investigate our research questions. Accordingly, we selected the DevOps webinars from 2017 to 2020, August (upon manually inspecting the available playlists for the most recent webinars) from the above source, i.e., 426 entries.

\subsection{Data extraction}
We carried out the following steps to extract data from the webinars.
We used a python API\footnote{https://pypi.org/project/youtube-extract/} to extract the metadata (Title, Link, View count, Duration, Date, Tags, and Description) of the 426 webinars. Firstly, the retrieved webinars were filtered based on the title (i.e., on whether any security-related aspect or DevSecOps topic was mentioned). We selected 95 webinars through this process. Then, we used the video description and when needed the video itself to apply the following inclusion (I1) and exclusion criteria (E1 and E2) to select 31 webinars for the analysis of our study.

\begin{itemize}
    \item [I1] A key focus is on including security tools into the DevOps workflow.
    \item [E1] The Webinar is addressing one narrow stage or technological component (e.g., databases) in the overall DevOps workflow.  
    \item [E2] Most of the webinar (e.g., more than half the duration) is centered around discussing features of commercial products.
\end{itemize}

\subsection{Data analysis}

We used the following steps of the thematic analysis \cite{braun2006using, cruzes2011recommended} method for the analysis stage of our study.

\textit{Familiarizing with data}: The first author had been following the webinars for one year. Therefore, the structure and typical nature of a webinar were known to him. The overview of the selected webinars (i.e., Title, URL, upload date, tags, and webinar description) was included in Excel sheets and shared among the other authors. 

\textit{Generating initial codes}: 
We obtained the transcripts of all the selected webinars using a Python API\footnote{https://pypi.org/project/youtube-transcript-api/}. We imported these transcripts into NVivo12, a qualitative data analysis software, to generate the \textit{initial codes}. A \textit{code} (a phrase that summaries the key points \cite{hoda2012developing}) was assigned to data segments from the transcripts which related to our research questions.  

We coded on the transcript of each webinar while watching the video of the relevant webinar on YouTube. Due to the limitations of the API, most of the technical terms and certain words were inaccurately transcribed. Further, punctuation marks such as full stops for the sentences were missing. Therefore, we corrected the transcript only on the data extracted as codes, while cross-checking with the video. As this was an iterative process, some of the codes were deleted, merged or split.

\textit{Searching for themes}: In this step, we assigned the codes to potential themes (Figure \ref{fig:coding}A). We used a multi-layered coding structure in Nvivo for this task (Figure \ref{fig:coding}B).

\begin{figure*}[t]
    \centering
    \includegraphics[origin=c, width=.7\textwidth]{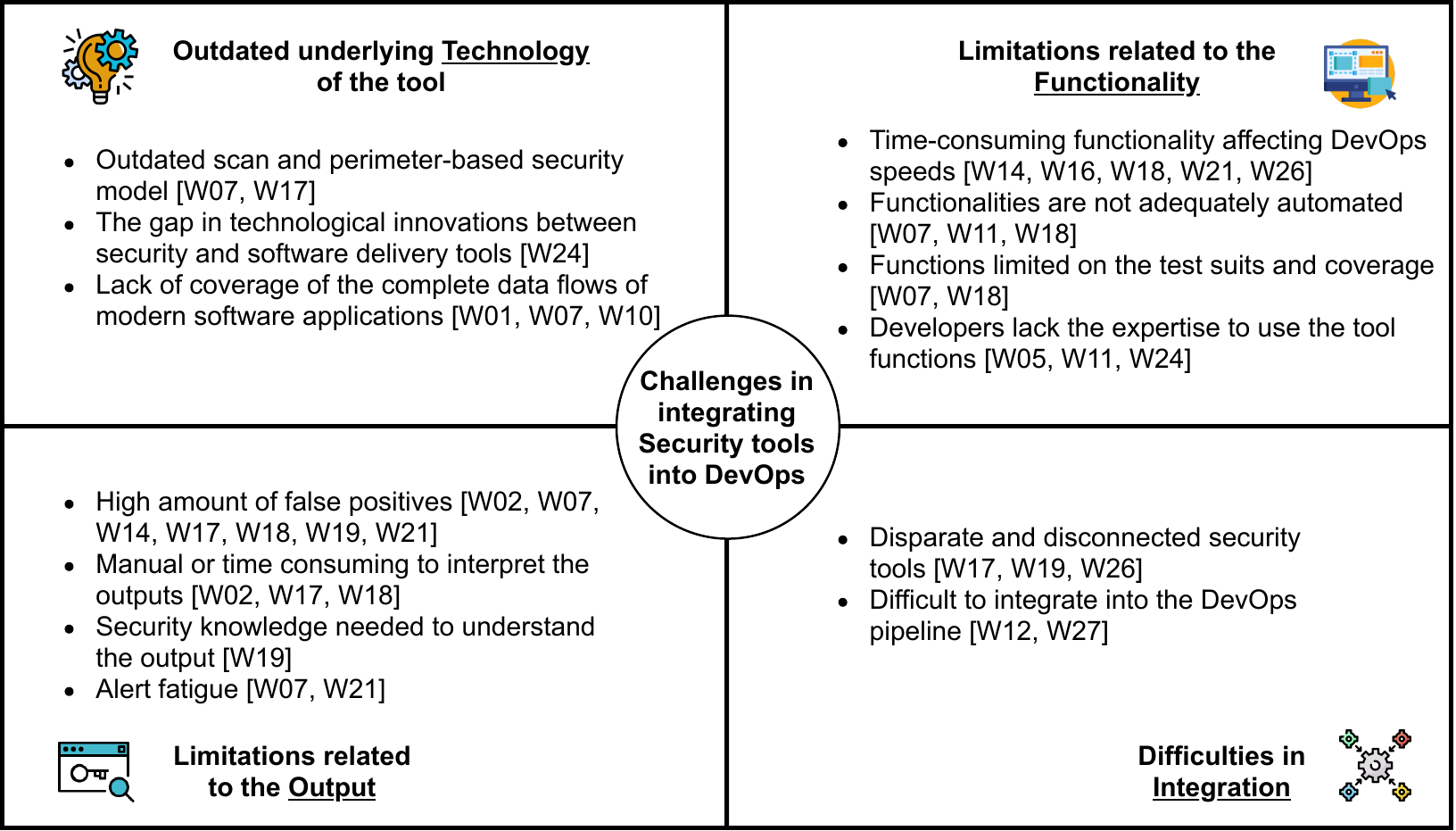}
    \vspace{-2mm}
    \caption[Caption for LOF]{Summary of RQ1: The challenges of integrating security tools into the DevOps workflow}
    \label{fig:RQ1}
\end{figure*}

\textit{Reviewing themes}: We iteratively modified the themes (e.g., reassigned codes among themes) in this stage. We created two \textit{higher-level categories} for the themes of the second research question.

\textit{Defining and naming themes}: We finalized the names of the themes in this stage. 

\textit{Producing report}: We report the results for our research questions in Section 4. For each research question, we present a summary figure that details the themes and key points that formed the codes. 

We carefully reviewed each step of the above process and held regular (e.g., weekly) meetings among ourselves to discuss the outputs. Figure {\ref{fig:method}} presents the summary of the method in our study.

\section{Results}

\subsection{Challenges of integrating security tools into the DevOps workflow}
We created four themes to classify the results of this research question (Figure \ref{fig:RQ1}). The following sections detail each of these themes.

\subsubsection{Limitations of the underlying technology of the security tools}
Our analysis has revealed that the limitations of the outdated underlying technology of the security tools are a significant cause for the disconnect between DevOps and security. Practitioners argue that there is an \textit{innovation gap} in the technologies between how we deliver software and produce secure outputs [W24]. For example, there have been numerous tools developed to improve how we build and deliver software in the past couple of years (e.g., Continuous Integration (CI)/ Continuous Deployment (CD) systems, containers, microservices). However, tools targeting security have only seen gradual technological advances (e.g., static and dynamic application security testing) [W24].  Most established security tools are also based on a scan and perimeter-based software security model [W17]. Practitioners believe that this is an outdated model, not suited for modern software development needs:

\textit{"The state of the practice for application security is based on a 15 year old scan and perimeter based software security model. This is built for the pre-DevOps, pre-cloud, pre-microservices era. So it's no wonder that it cannot keep pace with the needs of DevOps."} \textbf{[W17]}

Understanding the data flows in the application is critical for security assessment [W01]. However, another technological limitation of traditional security tools is that they lack coverage of the complete data flows related to modern architectures that are popular among DevOps teams:

\textit{"I'm not aware of a static analysis tool on the market that can stitch data flow together across a bunch of lambda functions or across a bunch of microservices. And the net effect of that is that you don't get a full picture of what the application is doing. And you may miss things"} \textbf{[W01]}

Tools such as Web Application Firewall (WAF) also lack the understanding of data flows and sensitive data classifications [W07]. For example, WAFs are installed at the edge of the application domain and only understand the inputs applied. However, if the application is communicating internally with components (e.g., microservices), these tools have a limited understanding of that communication [W07]. 

\begin{tcolorbox}[left=1pt, top=1pt, right=1pt, bottom=1pt]

\textbf{Ch1}: Due to the outdated underlying technologies used in security tools, they are unable to cater to modern software development methods and software delivery tools.

\end{tcolorbox}

\subsubsection{Limitations related to the security tool functionalities}
Limitations of the security tools' functionalities and their effect on DevOps were a key concern among practitioners. For example, the time-consuming functionalities of traditional security tools were a widely discussed drawback [W14, W16, W18, W21, W26]. In the case of SAST, conducting a full analysis (i.e., scanning the complete code base) or a deep scan takes a substantial amount of time [W18]. This is a limitation which affects DevOps, as pointed out by practitioners:

\textit{"For every release of the software created, it is critical to identify vulnerabilities at high precision in minutes. Because if a typical DevOps deployment pipeline comprises of compiling and deploying to production in minutes, we cannot be running static scans for hours."} \textbf{[W21]}

For DAST, the situation is even more critical, as practitioners note that it takes more time than SAST in a practical setting [W18].  This especially relates to dynamic tests such as fuzz testing [W18].

Practitioners further note that the traditional security tools have not been adequately automated [W11]. Therefore, substantial manual effort is needed to run these tools [W07, W18]. One example of this is the significant amount of manual tuning required for most traditional security testing tools [W07]. Every time new threats are identified (e.g., zero-day), vulnerability rules have to be manually created and updated in the databases of most of these tools [W07].  

Tools such as DAST depend on the effectiveness of the test suites. Nevertheless, developers create test suites to test the business workflow of an application, not necessarily the security efficiency [W07]. Furthermore, achieving complete security testing coverage using these tools is challenging, with the DevOps time restrictions [W18].

Lastly, our analysis has revealed that many developers lack the required expertise to use most of the available security tools [W05, W11, W24]. 

\textit{"There's a lot of skill that it takes to operate some of the web application scanning tools."} \textbf{[W24]}

Practitioners note that developers and security engineers have very different domain expertise [W05]. Hence, developers often do not have the necessary skills to use security tool functions. 

\begin{tcolorbox}[left=1pt, top=1pt, right=1pt, bottom=1pt]

\textbf{Ch2}: The functionalities of many security tools are not adequately automated or carry limitations that hinder their adoption in DevOps. Most developers also lack the required security skills to use these tools effectively.

\end{tcolorbox}

\subsubsection{Difficulties related to the tools' outputs}

Our data analysis has revealed that practitioners frequently complain about the high amounts of false positives in the security tools' outputs (e.g., SAST, DAST, WAF, RASP) [W07, W18, W19, W21]. In this context, a false positive is when an application's security scanner indicates that there is a vulnerability, but in reality, it is a false alarm [W14]. While this is noted as a drawback even in the traditional setup, practitioners note that false positives have a more severe effect in a rapid deployment environment:

\textit{"Because of the high false positives, the distrust begins to essentially manifest between security and engineering. Because for security, they want most of the issues triage and addressed. For engineering, they want to deploy their software at high velocity [..]"} \textbf{[W07]}
 
Practitioners also state that if a tool output contains a high amount of false positives, engineers will assume that all remaining messages are also inaccurate [W14]. Thus, an engineer is most likely to ignore the remaining output [W14].  

Subsequently, this issue results in a very high operational cost, as a security engineer needs to identify the true positives. This is because many developers are unable to understand the output of security tools and provide remedies [W19]. Therefore, security engineers are required to distinguish exploitable issues from non-exploitable ones manually [W18].  

To overcome individual security tool limitations, practitioners recommend using multiple tools in the pipeline. However, with the development or security teams requiring to use several different application security tools (e.g., SAST, DAST, WAF, RASP), alert fatigue becomes a problem [W21]. Alert fatigue occurs when an overwhelming number of alerts desensitizes the person who is responsible for addressing them \cite{Atlassian}. This could lead to ignored or missed true positive alerts. 

\begin{tcolorbox}[left=1pt, top=1pt, right=1pt, bottom=1pt]

\textbf{Ch3}: Limitations related to the outputs of traditional security tools hinder rapid deployments as manual and time-consuming tasks that require security expertise are needed to address these issues.
\end{tcolorbox}

\subsubsection{Difficulties in tools integration}
Integration of tools plays an important role in DevOps [W12]. Here, the software delivery tools form a \textit{pipeline} for continuous and rapid delivery. Therefore, security tools need to integrate into this pipeline. However, practitioners state that smoothly integrating traditional security tools into a DevOps pipeline is very difficult [W27]. They also note that integrated security tools are needed in an environment where multiple such tools are used. However, lack of integrated security tools was mentioned as a challenge in several webinars [W17, W19, W26]:

\textit{"If you look at the state of the practice that people have employed to date for application security [..] you get what we call this legacy tool quagmire. You have a set of disparate, static, disconnected tools that taken together are largely inaccurate"} \textbf{[W17]}

This issue leads to such tools used in isolation. Practitioners stated that each tool would then only uncover a subset of the total vulnerabilities present in a given software [W19]. 

\begin{tcolorbox}[left=1pt, top=1pt, right=1pt, bottom=1pt]

\textbf{Ch4}: Integration of tools into a pipeline is a key technical requirement in DevOps. Yet, most security tools are difficult to integrate into this pipeline and with other security tools.

\end{tcolorbox}

\subsection{Practitioner recommendations to integrate security tools in a DevOps workflow}

\begin{figure*}[t]
    \centering
    \includegraphics[origin=c, width=.75\textwidth]{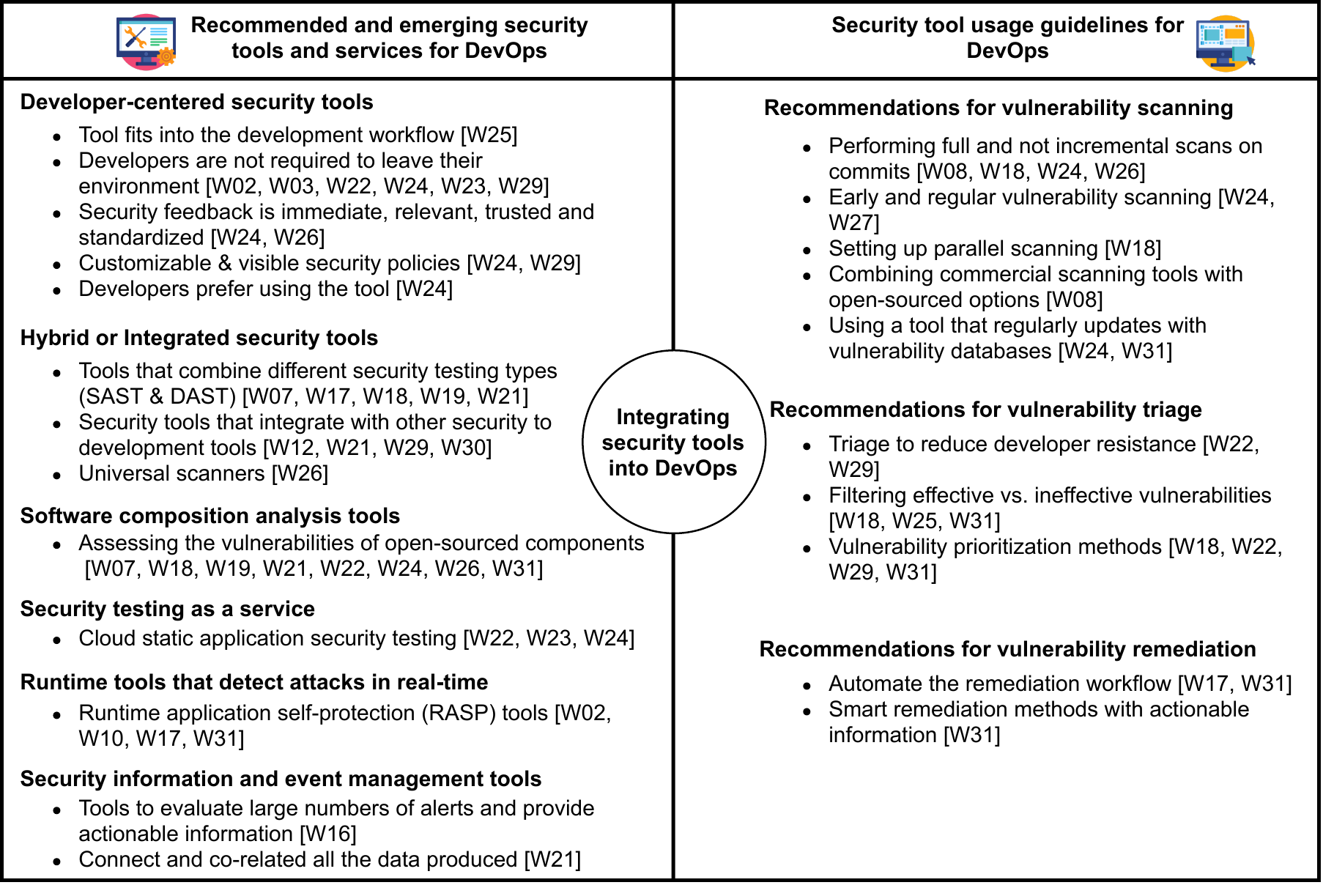}
    \vspace{-2mm}
    \caption{Summary for RQ2: Integrating security tools into DevOps}
    \label{fig:RQ2}
\end{figure*}

The following sections present the two higher-order categories and details of the themes of the second research question (Figure \ref{fig:RQ2}).

\subsubsection{Security tools and services that target DevOps}
In the webinar discussions, practitioners recommended certain emerging tools or services that are more suited for DevOps. These tools either target specific DevOps practices (e.g., shift-left and developer centred security) or address limitations of traditional tools. Therefore, these tools support better integration into the DevOps workflow.

\paragraph{\textbf{Developer-centered security tools}}

Our analysis has revealed that most application security tools are catered for application security professionals [W24]. However, in DevOps, developers are required to engage in security tasks. Therefore, practitioners advocate for \textit{developer-centered security tools} for this domain. We extracted some key attributes for such tools from the webinar discussions. 

Firstly, developers need security tools that they prefer to use [W24] and fit into their workflow [W25]. A frequently discussed recommendation was that developers should not be required to leave their development environment to engage with the security tools [W02, W03, W22, W24, W23, W29].

\textit{"Bring the tools where the developer lives. Don't expect them to be logging into other systems."} \textbf{[W29]}

An example of this is static analysis tools that integrate into the Integrated Development Environment (IDE) [W02]. Such tools enable developers to run scans and discover vulnerabilities as they code. Similar to compiler errors, these tools can point out security-related errors while developers code [W03, W24]. Such feedback would enable developers to fix vulnerabilities immediately in a process that is familiar to them [W24]. However, the integrated static analysis tools require an efficient code scanning functionality with immediate feedback [W24]. Similar to speed, accuracy is another important requirement: 

\textit{"Accuracy is important primarily because we got to earn those developers' trust [..] And if you can deliver the accuracy so that people stop arguing about whether or not the results are real [..]"} \textbf{[W24]}

Practitioners also discussed the need for security feedback following a standardized format (e.g., Static Analysis Results Interchange Format (SARIF))[W26]. If tools use such standards, developers will find it easier to interpret the feedback, especially when multiple security tools are utilized.

Lastly, visible security policies are another requirement for developer-centered security tools that we have captured in our study [W24, W29]. These policies need to be customizable (e.g., \textit{"what level of vulnerabilities an application should have?"} [W24]) and easy to implement, particularly for developers without specialized security training. Having visible policies would also reduce the developer resistance to address the outputs of security tools.

\paragraph{\textbf{Hybrid or integrated security tools}}

Hybrid tools are an emerging type of tool suited for DevOps that practitioners discussed in the analyzed webinars. Tools that combine application security testing technologies (e.g., SAST, DAST, and RASP) are examples of such tools [W07, W21]. The resulting hybrid tool would consist of the strengths of each of the combined technologies [W07]. For example, Interactive Application Security Testing tools (IAST) combine the strengths of both SAST and DAST tools [W19, W18, W19]. This technique detects vulnerabilities in both custom code and libraries during normal use of the application [W17]. IAST contains agents that deploy sensors inside an application run-time environment, which continuously analyzes the application interactions to identify vulnerabilities in real-time. This approach enables a security tool to minimize false positives and improve accuracy [W17]. Further, certain critical vulnerabilities that only occur during run-time can be detected by IAST [W19]. Therefore, this is an emerging type of tool that practitioners recommended for DevOps workflows:

\textit{"How do we prioritize critical vulnerabilities that will help minimize the development team's effort in triaging and remediation? This is where IAST comes in [..] it is a neat next-generation solution that can help build in continuous security testing in CI/CD"}\textbf{ [W19]}

Despite the availability of many security tools, it is difficult practically to interact with several tools at once in a rapid deployment environment. Therefore, practitioners recommend security tools that can be integrated with other development or security tools and platforms (e.g., integrates with bug tracking or ticket management tools)  [W12, W21, W29, W30]. Such integrated tools would also satisfy the practitioners' requirements for \textit{Universal security scanners}: 

\textit{"We wanted (product name) to be a universal security scanner. We were tired of using scanners for different languages, different frameworks, different purposes (so) we wanted to build that universal layer that abstract (the) security scanning process."} \textbf{[W26]}

Ultimately, these integrated tools would be easier to be seamlessly added to a CI/CD pipeline in the DevOps workflow.

\paragraph{\textbf{Software composition analysis tools}}
Due to the heavy utilization of free and open-source software (FOSS) components in DevOps projects, practitioners discussed the importance of assessing the security of such components:

\textit{"Vulnerabilities might not just lie in the code that your engineers are creating. Vulnerabilities might lie in the open-source utilized by your custom code. So it is critical to examine your application in its entirety."} \textbf{[W07]}

A specialized technology that is used to detect vulnerabilities in FOSS is software composition analysis (SCA) tools. Practitioners discussed the importance of having SCA in the DevOps pipeline as a separate tool or integrated as a component to an overall code scanning solution  [W07, W18, W21, W24, W26, W31]. Such tools need to be configured based on the needs (e.g., delivery frequency, FOSS components used) of the DevOps pipeline.

\paragraph{\textbf{Security testing as a service}}

To overcome the difficulties of using the existing application security tools in a DevOps environment, practitioners presented the option of acquiring the tool functions as a service [W22, W23, W24]. For example, static analysis is now delivered in the cloud by service providers, and practitioners discussed the trade-offs of this option: 

\textit{"The trade-off between using an on-premise tool, which is going to have significant hardware limits, and by large doesn't have the accuracy [..] versus doing something in the cloud that is enabling you to reduce your MTTR so dramatically, and fix vulnerabilities in the same sprint [..]"} \textbf{[W24]}

According to the above quote, the benefits of using a cloud service outweigh the potential security concerns (e.g., code being exposed to an external party) [W24]. A key advantage discussed is the substantial reduction of the MTTR (mean time to recovery or mean time to restore) [W24] and the ability to rectify vulnerabilities rapidly, particularly without in-house security experts. Therefore, based on these discussions, the security as a service model would suit the rapid deployment needs of DevOps [W22, W23, W24].

\paragraph{\textbf{Run-time tools that detect attacks in real-time}} Run-time application self-protection (RASP) are another type of tool recommended for DevOps [W02, W10, W17, W31]. Practitioners noted that these tools provide accurate and more fine-grained diagnostic details (compared with similar older technologies) [W17]. Similar to IAST, RASP is embedded within an application and is continually scanning for threats and malicious attempts [W17]. Practitioners discussed the benefits of RASP as follows:

\textit{"We spoke about like 100s or 1000s of vulnerabilities in an average application, probably even if we use the right tools, we will not remediate all over them. So still, RASP, as a security technology, is sort of another layer of protection, but maybe sometimes it will be your only layer of protection."} \textbf{[W31]}.

\paragraph{\textbf{Security information and event management tools (SIEM)}} 
Developers experience \textit{alert fatigue} due to working with many tools in a DevOps pipeline [Ch3]. To overcome this issue, practitioners recommended using Security Information and Event Management (SIEM) tools in this context [W16, W21]. A SIEM is a platform or set of tools that provide a holistic view of security events or incidents. In the analyzed webinars, practitioners discussed how a SIEM could be used to address the large number of alerts generated by various tools used in the pipeline:

\textit{"It is the onus of the SIEM and its correlation engine to connect all the data produced or alerts produced by these systems to provide enough data to prioritize which vulnerabilities need to be addressed."} \textbf{[W21]}.

As noted, such a solution makes it easier for developers to select which security issues to prioritize amid many alerts. Practitioners also stated that a SIEM could help deal with different disconnected tools creating alerts used in different phases of the pipeline [W21].

\begin{tcolorbox}[left=1pt, top=1pt, right=1pt, bottom=1pt]

\textbf{T1}: As developers are expected to engage in security tasks in DevOps, Developer-centered security tools are a key requirement in this domain.

\textbf{T2}: Hybrid and integrated security tools are more suited for DevOps as they combine the advantages of security testing technologies and avoid some disadvantages. 

\textbf{T3}: Due to the heavy utilization of FOSS in DevOps projects, using an SCA tool in the pipeline is highly recommended.

\textbf{T4}: The security as a service model overcomes many drawbacks of traditional security tools that limit its adoption in DevOps.

\textbf{T5}: RASP, compared with traditional firewalls, provide more accurate and fine-grained diagnostic details.

\textbf{T6}: SIEM tools are beneficial in DevOps for dealing with the large amounts of alerts generated by disconnected security tools.

\end{tcolorbox}

\subsubsection{Guidelines for using security tools and assessing their outputs in DevOps}

This section presents the result of our analysis highlighting practitioners' recommendations for using and assessing the outputs (e.g., vulnerability remediation) of security tools in a DevOps workflow. These practices also aimed to diminish the limitations of security tools or optimize the DevOps workflow.

\paragraph{\textbf{Recommendations for vulnerability scanning in DevOps.}}
Vulnerability scanning of source code using security tools (e.g., SAST) is common practice in the industry. However, the existing static analysis tools suffer from drawbacks, particularly when used in a rapid deployment environment [Ch1]. Therefore, more efficient ways to perform this task using security tools are needed for DevOps. Therefore, we have derived several recommendations for vulnerability scanning in a DevOps setting from the webinars.

In DevOps, regular commits to the mainline code branch are performed by developers. In such a scenario, practitioners recommend conducting security scanning at every pull request or commit [W08, W18, W24, W27]. This would enable developers to address vulnerabilities or other security issues in the same sprint they are found. Compared with the traditional practices, where the application security team does the scanning at a later stage of a project, this approach deals with vulnerabilities far more efficiently. 

In addition, when conducting vulnerability scanning, the developers are expected to perform a \textit{full scan} (as opposed to incremental scans) [W08, W18, W24, W26]. A full scan is required to assess how the new code interacts with the rest of the codebase, which would be necessary to uncover vulnerabilities: 

\textit{"If you think about vulnerabilities and the relationship with data flow, you can't really get the same level of comprehensiveness, if you're only doing an incremental scan, you really need to understand how the new code interacts with the rest of the codebase. And that requires a full scan."} \textbf{[W24]}

However, due to the limitations of the traditional tools in performing full scans (e.g., Ch2), practitioners recommend creating an effective scanning strategy for rapid deployment environments. For example, regarding dynamic scanning, practitioners discuss performing the scan in parallel in the pipeline [W18]: 

\textit{"Run dynamic testing in the parallel mode, so they don't stop the pipeline. If there is anything identified, like vulnerability or severe security issue, but they create the tickets [..] the pipeline is not stopped"} \textbf{[W18]}

Another best practice recommended by practitioners is to combine scanning tools to minimize the limitations of the scanning technologies (e.g., false positives):

\textit{"The good organizations always use two to three different scanners. If for nothing else, just to see the difference (of the result)."} \textbf{[W08]}

For example, if you are using a commercial scanning tool, you could also use a FOSS tool, particularly as many adversaries use such tools for malicious activities [W08].

Practitioners also discussed some properties to be considered in selecting a vulnerability scanning tool. For example, the tool should be scanning code for the most frequently occurring security issues, such as the Open Web Application Security Project (OWASP) top 10 [W24]. Therefore, a security tool should conduct regular updates with common vulnerability databases [W24]. In addition, the tool should also assess data leakage and business logic flaws that are unique to a particular codebase [W24]. Such issues can be hard to detect but commonly occurring, especially in a rapid deployment environment. 

\paragraph{\textbf{Effective vulnerability triage.}}
Vulnerability triage is the process of identifying and prioritizing critical vulnerabilities that will help minimize a development team's effort in remediation [W19]. This process plays a vital role in DevOps due to the rapid deployment needs and many alerts or warnings typically generated by security tools. It is practically difficult for developers to address all warnings or alerts produced by the security tools with the needs of DevOps [W29]. 

A key recommendation in triaging is to use an established vulnerability metric such as the Common Vulnerability Scoring System (CVSS) to score and rank vulnerabilities [W22]. By doing so, vulnerabilities can be addressed based on the level of severity. Another suggestion regarding triaging is to differentiate between effective and ineffective vulnerabilities, as stated below: 

\textit{"So if you have a vulnerable method and a library you're using, that's one thing, (but) if you're calling that vulnerable method, that's a much higher likelihood of attack surface. And that's a huge deal."} \textbf{[W25]}

Practitioner-based surveys have reported that only 30\% of the total amount of results from a scan contain effective vulnerabilities [W25]. Therefore, in a DevOps environment, these vulnerabilities need to be prioritized in the remediation process. 

Finally, the importance of giving the triage processes visibility and transparency is discussed [W04]. Better visibility to this process would enable all the relevant stakeholders to provide critical input, which would aid in the prioritization of severe issues.

A positive outcome of the vulnerability triage is to reduce the developer resistance for vulnerability remediation [W22, W29]. Due to large numbers of false positives generated by security tools, developers tend to oppose remediation tasks based on the results of these tools [W29]. However, if vulnerability triage processes are able to identify potential vulnerabilities that could lead to substantial damages, this resistance could be minimized. 

\paragraph{\textbf{Recommendations for vulnerability remediation}}

Vulnerability remediation is the process where developers address the security issues resulting from the vulnerability triage [W17, W31]. Developers should be given accurate, actionable information with the maximum impact for this process as there is limited time in DevOps [W31]. Also, due to the rapid delivery needs of DevOps, practitioners state that the vulnerability remediation workflow also needs to be automated to a greater extent [W17, W31]:

\textit{"So only if the vulnerability is high, I want to actually open a pull request to remediate it automatically, or based on the CVSS score. Or [..] whenever there is a new version for the open-source, I want to automatically upgrade and remediate any potential vulnerabilities."} \textbf{[W31]}

However, practitioners note that for an automated remediation workflow to be successful, many automated tests based on clearly defined roles and policies need to be in place [W31].  

\begin{tcolorbox}[left=1pt, top=1pt, right=1pt, bottom=1pt]

    \textbf{G1}: A vulnerability scanning strategy that suits a rapid deployment environment needs to be formed to minimize the drawbacks of security testing tools.
    
    \textbf{G2}: A visible vulnerability triage process that prioritizes and differentiates between effective and ineffective vulnerabilities is important in this domain to reduce developer resistance to vulnerability remediation.
    
    \textbf{G3}: Automated vulnerability remediation methods are suited for this domain due to the fast deployment needs in DevOps.

\end{tcolorbox}

\section{Discussion}
\subsection{The need for new generations of tools}

Practitioners commonly agree that the tools based on traditional and established technologies are difficult to be integrated into the DevOps workflow [Ch4]. For example, SAST contains some inherent drawbacks in the technology, making it hard to scale into DevOps \cite{rajapakse2021challenges}. This is also true for several other types of security testing methods, such as DAST (e.g., fuzz testing) or penetration testing \cite{rajapakse2021challenges}. Consequently, developers face difficulties in performing adequate security testing due to the limitations of these technologies, considering the requirements of DevOps.

\begin{figure*}[t]
    \centering
    \includegraphics[origin=c, width=.95\textwidth]{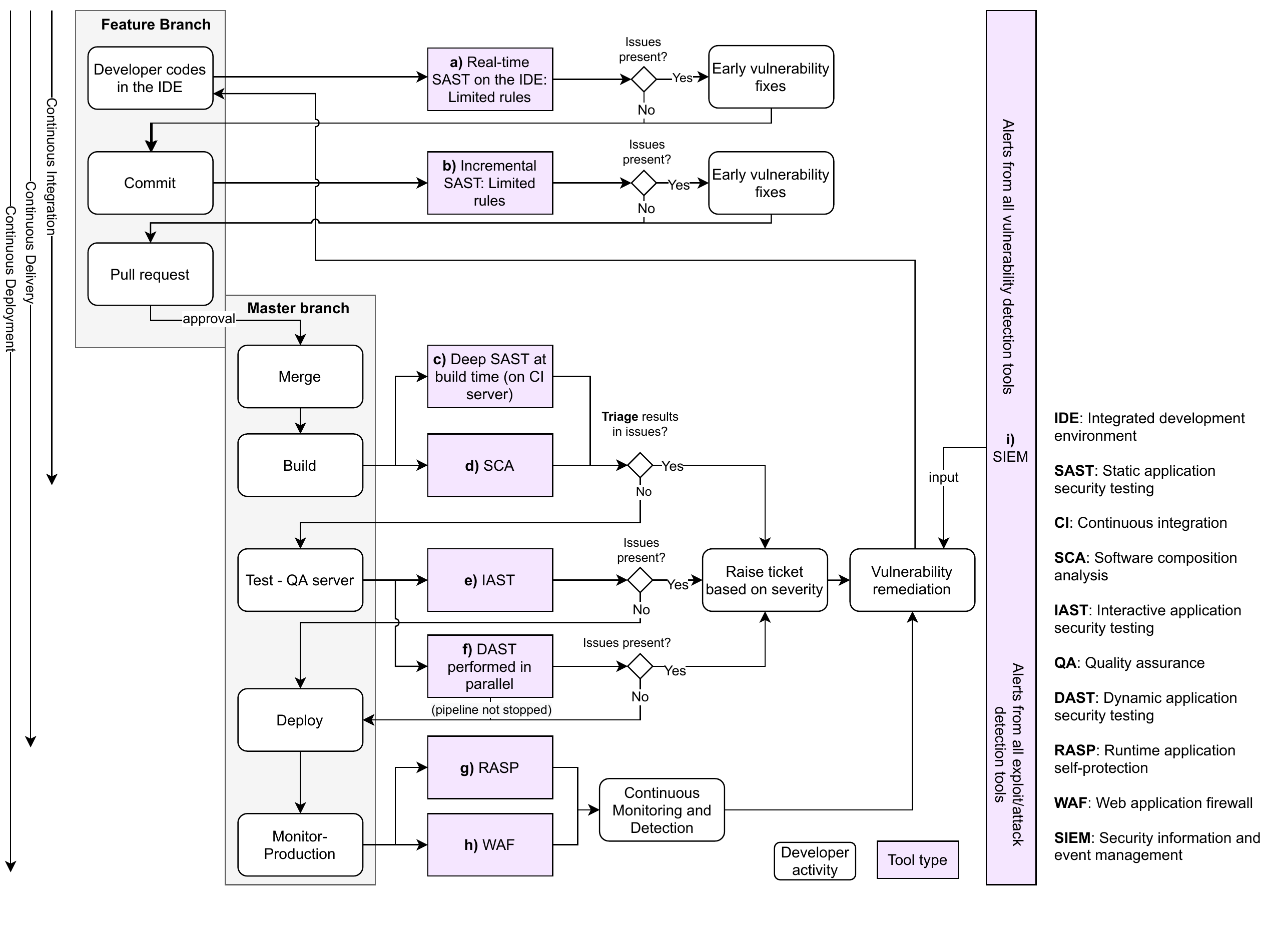}
     \vspace{-10mm}
    \caption{DevOps workflow synthesized from the webinars [W02, W04, W09, W18, W19, W21, W24]}
    \label{fig:workflow}
\end{figure*}

Given the results of our analysis, we highlight the existence of a technological \text{innovation gap} between application security and software delivery tools. In recent times, software delivery tools and methods have undergone drastic improvements, adapting to the new software development paradigms such as DevOps \cite{kersten2018cambrian}. However, the currently established application security technologies are still targeting the traditional software development cycle \cite{Checkmarx2020}. In the traditional setting, after the development activities conclude, the code is handed over to testing teams (i.e., activities follow sequentially). SAST and DAST tools work well in this model, where security engineers are able to perform comprehensive and time-consuming security tests. The code is deployed after the results of such tests are deemed satisfactory. However, developers face many challenges in using the same tools in DevOps, where the development workflow and the goals are vastly different. 

Our analysis revealed several recommendations or guidelines for integrating security tools into the DevOps workflow while overcoming certain tool limitations. However, practitioners note that there are many challenges yet to be adequately addressed. For example, given the SAST limitations, performing the full scan on every software change (e.g., commit) is still challenging if the deployment frequency is high.

Further, our study captured some emerging security tools or technologies discussed by practitioners (e.g., IAST and RASP). However, both IAST and RASP are still relatively immature technologies and, therefore, yet to arrive at widespread usage in the industry \cite{Lemos}. While these technologies are a step forward from SAST and DAST, they have not entirely resolved the security testing challenges practitioners face in a DevOps setting. Furthermore, these tools have their own drawbacks. For example, IAST and RASP can slow down the operations of an application \cite{PositiveTechnologies2019}. Therefore, more work is needed towards developing the next generation of security tools that are suited for modern software development paradigms.

\subsection{Integrating security tools in the DevOps workflow}

We present an aggregated workflow based on the synthesis of the discussions in several analyzed webinars [W02, W04, W09, W18, W19, W21, W24] in Figure \ref{fig:workflow}. This workflow integrates most of the security tools and the tool usage recommendations of our study. We describe each tool integration point into the workflow below.

\textbf{a)} A SAST tool needs to be integrated into the IDE to enable static scanning while the developers are engaged in the development tasks (i.e., \textit{real-time SAST}). A limited number of rules should be tested on source code that developers are working on to obtain the results fast (e.g., configured to detect most frequently occurring vulnerabilities) [G1]. If vulnerabilities are present, developers can rectify the issues before committing the code [T1].

\textbf{b)} SAST on the commit should be performed using a tool integrated into the source code management system. A pre-defined set of rules should be tested only on the commit code to provide fast feedback (i.e., \textit{incremental SAST}) [G1]. If vulnerabilities are present, developers can rectify them and push a commit again [T1].

\textbf{c)} SAST needs to be performed at build time, using a SAST tool that is integrated into the build server. The scan needs to be run based on a comprehensive rule-set on the entire integrated code base (\textit{deep SAST}) [G1]. If security issues (e.g., vulnerabilities) are uncovered in the vulnerability triage [G2], the build should be broken and a ticket raised. Such issues should then be subjected to the vulnerability remediation process [G3].

\textbf{d)} Vulnerabilities of FOSS components should be identified at build time, using an SCA tool integrated into the CI or build server [T3]. This tool should be run on the entire integrated codebase to include all components (e.g., external libraries). If issues are present, the same SAST remediation process needs to be followed [G3].

\textbf{e)} Using an IAST tool in the quality or test server, source code should be analyzed for security vulnerabilities while the software is being run [T2]. If issues are detected, a ticket should be raised based on the severity. These tickets then need to be subjected to the vulnerability remediation process.

\textbf{f)} DAST needs to be performed in parallel in the quality server [G1]. The pipeline should not be stopped until DAST (i.e., all test cases) is complete. If issues are uncovered, the same vulnerability remediation process as IAST should be followed. \textit{All vulnerability related alerts from the above tools needs to be fed to a SIEM.}

\textbf{g)} A RASP tool should be deployed in the run-time environment to continuously and actively monitor deployed software for attacks [T5]. 

\textbf{h)} A WAF should be deployed in production to provide continuous protection/monitoring against external attacks. \textit{All attacks/exploits related alerts from WAF and RASP also need to be fed to a SIEM.}

\textbf{i)} A SIEM solution is needed to capture, evaluate and prioritize alerts generated by all tools across a pipeline [T6]. A SIEM tool would offer actionable information to developers (e.g., typically after a review) in the remediation process.

\subsection{Implications for practice and research}

Our results provide a synthesized set of information on security tool selection and usage in a DevOps environment. We discuss the limitations of traditional tools and some usage guidelines on reducing negative effects. We also present the state of the art tools currently emerging in the industry. Practitioners can consider adopting such tools in their DevOps projects, considering the advantages of these new technologies. Finally, we present a DevOps workflow that integrates both traditional and emerging security tools. Practitioners are able to adopt and modify this workflow depending on their context and needs.

We have pointed out the limitations of established security tools in a DevOps environment. Therefore, new generations of security tools are required in this context. This is an area for future research and development. We also see limited security and DevOps tool-related research on issues such as tool selection, providing tool support (e.g., error handling) and integration problems, even though there is high practitioner interest \cite{le2021large, zahedi2020mining, riungu2016devops, erich2017qualitative}. There is also a growing body of machine learning applications targeting the improvement of vulnerability assessment \cite{ghaffarian2017software, le2019automated, grieco2016toward}. However, more work is needed on how these advancements could be integrated into security tools to improve rapid deployment goals.

\section{Threat to Validity}

We have limited our search for webinars to one data source, i.e., DevOps.com. Therefore, our results are limited to the discussions carried out during the webinars organized by this platform. However, the facilitators of the webinars invited participants from a range of roles (e.g., developers, security engineers, DevOps advocates, solutions architects, product managers) and organizations. Further, it is one of the largest sources of DevOps content available.

In the filtering stage, all webinars with the relevant data may not have been included due to the limitations of the selection process (e.g., filtering based on title and webinar description). Further, the inclusion or exclusion of the webinars might be biased based on the research teams knowledge of the domain. The first author carried out the selection of the webinars for data collection after consultation with the co-authors. Then, the other authors reviewed the selected lists. The extraction process was also affected by the limitations of the transcription API (detailed in Section 3.3). 

The data analysis might also be affected by authors' bias and knowledge, as the first author conducted the coding of data. However, we have tried to reduce this threat by maintaining regular review meetings among all the authors to review the coding structure and verifying the results. We were also unable to validate some of the devised results (e.g., proposed workflow) of our study. Further, there is a threat related to the speakers of the selected webinars not being representative of the wider practitioner community. We plan to address these limitations of our study in future work.

\section{Conclusion}
As one of the main reasons why security is falling behind in DevOps, practitioners point to the gap in the technological innovations between security tools and software delivery technologies. As a result, most of the existing security tools have many drawbacks that hinder the rapid deployment cycles of DevOps. However, security vendors and open-source products are catching up to fulfil the needs of DevOps with the release of new generations of security tools such as IAST and RASP. These technologies point to the need for tools that understand more than one type of security testing technology.

For security to be successfully integrated into DevOps, merely using suitable security tools is not sufficient. Practitioners need to adopt suitable workflows which seamlessly integrate security tools. We have presented such a workflow derived based on our analysis of the webinars included in this study. We plan to further validate this workflow using an interview (small sample) and survey (larger sample) based study in future work.

Finally, according to our findings, we conclude that while security tools have started to cater to the rapid delivery needs of DevOps, there is an urgent need of allocating more attention and resources for developing and evaluating suitable security tools for DevOps. We hope that the findings from our study can provide useful insights for identifying the requirements and design options for the next generation of security tools for DevOps. 

\section{Acknowledgement}
The work has been supported by the Cyber Security Research Centre Limited whose activities are partially funded by the Australian Government’s Cooperative Research Centres Programme. We also thank the anonymous reviewers for their helpful feedback.

\bibliographystyle{ACM-Reference-Format}
\bibliography{main} 

\setcounter{secnumdepth}{0} 

\footnotesize
\begin{itemize}
\item [Icons] The free icons in Figure 3 and 4 are from  www.flaticon.com (Freepik, ddara)
\end{itemize}

\subsection{Selected Webinars of the study}
\footnotesize

\begin{itemize}
\item [W01] DevOpsTV. 2017. Protect Your Organization Against Known Security Defects. Retrieved from: \url{https://www.youtube.com/watch?v=hDAnvFlZOWw}

\item [W02] DevOpsTV. 2017. Take Control: Design a Complete DevSecOps Program. Retrieved from: \url{https://www.youtube.com/watch?v=gftkHoPQUIg}

\item [W03] DevOpsTV. 2017. Getting Started with Secure DevOps. Retrieved from: \url{https://www.youtube.com/watch?v=D3C318mDsjU}

\item [W04] DevOpsTV. 2018. Seven Deadly Saves To Security With Integrations. Retrieved from: \url{https://www.youtube.com/watch?v=1mFOy7z0NY0}

\item [W05] DevOpsTV. 2018. Shift Left Security The What, Why and How. Retrieved from: \url{https://www.youtube.com/watch?v=I8OSX4Kk97o}

\item [W06] DevOpsTV. 2018. Integrating Security into your Development Pipeline. Retrieved from: \url{https://www.youtube.com/watch?v=_6H_b8C8vJ8}

\item [W07] DevOpsTV. 2018. How-to Automate Application Security \& Keep Up with Modern CI-CD. Retrieved from: \url{https://www.youtube.com/watch?v=52FyxVgqy8M}

\item [W08] DevOpsTV. 2018. You Build It, You Secure It: Higher Velocity and Better Security with DevSecOps. Retrieved from: \url{https://www.youtube.com/watch?v=ZZ8GuSnHJQs}

\item [W09] DevOpsTV. 2018. This Year at RSA: Don’t Miss The Conversation on DevSecOps!. Retrieved from: \url{https://www.youtube.com/watch?v=t08vNQzSf7s}

\item [W10] DevOpsTV. 2017. From Good Code to Great Code: Why Developers Need to Own Application Security. Retrieved from: \url{https://www.youtube.com/watch?v=S9Uno7yyTMI}

\item [W11] DevOpsTV. 2018. Embrace DevSecOps and Enjoy a Significant Competitive Advantage!. Retrieved from: \url{https://www.youtube.com/watch?v=sv7vVYhxgrs}

\item [W12] DevOpsTV. 2018. DevOps Security: Build-Time Identification of Security Issues. Retrieved from: \url{https://www.youtube.com/watch?v=DfuYTRxP5Dg}

\item [W13] DevOpsTV. 2018. Shifting Left…AND Right to Ensure Full Application Security Coverage. Retrieved from: \url{https://www.youtube.com/watch?v=Ya6VX28TNGc}

\item [W14] DevOpsTV. 2019. Redefining CLI to Unify Security and DevOps. Retrieved from: \url{https://www.youtube.com/watch?v=8ST7g3NVmfg}

\item [W15] DevOpsTV. 2019. How Cloud-Native has Changed Application Security. Retrieved from: \url{https://www.youtube.com/watch?v=IzYSwIJUby0}

\item [W16] DevOpsTV. 2019. The Impact of Digital Transformation on Enterprise Security. Retrieved from: \url{https://www.youtube.com/watch?v=kwXUJR_s0NI}

\item [W17] DevOpsTV. 2019. Embracing DevSecOps with Embedded Application Security. Retrieved from: \url{https://www.youtube.com/watch?v=JzGBCmIaLAo}

\item [W18] DevOpsTV. 2019. Security in CI CD Pipelines: Tips for DevOps Engineers. Retrieved from: \url{https://www.youtube.com/watch?v=S7TfXEyhLck}

\item [W19] DevOpsTV. 2019. Bridging the Security Testing Gap in Your CI/CD Pipeline. Retrieved from: \url{https://www.youtube.com/watch?v=aWWjvy9UpfY}

\item [W20] DevOpsTV. 2019. Building Resilience into Your DevOps Environment. Retrieved from: \url{https://www.youtube.com/watch?v=-JjERAP9k5Y}

\item [W21] DevOpsTV. 2019. Inserting Security into DevOps Pipelines the Fast Way. Retrieved from: \url{https://www.youtube.com/watch?v=waWS9jfW6Ao}

\item [W22] DevOpsTV. 2019. DevOps Tools: DevSecOps Tools Worth Knowing. Retrieved from: \url{https://www.youtube.com/watch?v=qZX--tdwY6A}

\item [W23] DevOpsTV. 2019. Continuous Compliance and DevSecOps in Times of GDPR, HIPAA and SOX. Retrieved from: \url{https://www.youtube.com/watch?v=ZSj46aIbkOI}

\item [W24] DevOpsTV. 2020. Making Security More Efficient for Developers. Retrieved from: \url{https://www.youtube.com/watch?v=4q-dIeI20RQ}

\item [W25] DevOpsTV. 2020. From Zero to DevSecOps How to Implement Security at the Speed of DevOps. Retrieved from: \url{https://www.youtube.com/watch?v=oS-SIkO_8cQ}

\item [W26] DevOpsTV. 2020. An Open Source DevSecOps Platform for Securing Code \& Dependencies. Retrieved from: \url{https://www.youtube.com/watch?v=MVa9smsCI5Q}

\item [W27] DevOpsTV. 2020. Do You Trust Your DevSecOps Pipeline. Retrieved from: \url{https://www.youtube.com/watch?v=X3TMzVKnA1A}

\item [W28] DevOpsTV. 2020. Beyond the Top 10 Finding Business Logic Flaws, Data Leakage and Hard Coded Secrets in Development. Retrieved from: \url{https://www.youtube.com/watch?v=1dLGQ5FE9CY}

\item [W29] DevOpsTV. 2020. Whose Vulnerability Is It Anyway?. Retrieved from: \url{https://www.youtube.com/watch?v=wXJSUwyUZpA}

\item [W30] DevOpsTV. 2020. 7 Techniques for Ramping Your DevSecOps Program Quickly. Retrieved from: \url{https://www.youtube.com/watch?v=tIGtS63Wj84}

\item [W31] DevOpsTV. 2020. DevSecOps: Closing the Loop from Detection to Remediation. Retrieved from: \url{https://www.youtube.com/watch?v=0ZubO-oty9s}

\end{itemize}

\end{document}